\documentclass[twocolumn,10pt,journal]{IEEEtran}

\usepackage{graphicx}  
\newif\ifpdf
\ifx\pdfoutput\undefined
\pdffalse 
\DeclareGraphicsExtensions{.eps} \else
\pdfoutput=1 
\pdftrue \DeclareGraphicsExtensions{.pdf,.jpg,.png} \fi

\begin{document}

\title{ Analytical Model of Wireless Cell with Superposition Coding 
}

%

\author{
\IEEEauthorblockN{
Jean-Marc Kelif$^1$,\thanks{$^1$Jean-Marc Kelif is with Orange Labs, France.
E-mail: jeanmarc.kelif@orange.com} 
Jean-Marie Gorce$^2$,\thanks{$^2$Jean-Marie Gorce is with Univ Lyon, INSA Lyon, Inria, CITI, 
Villeurbanne, France. E-mail: jean-marie.gorce@insa-lyon.fr} 
Azeddine Gati$^3$\thanks{$^3$Azeddine Gati is with Orange Labs, France. E-mail: azeddine.gati@orange.com} 
}}



%


\maketitle

\begin{abstract}

Focusing on the downlink, we consider a base station (BS) and the cell it covers. 
The Superposition Coding (SC), also referred to as Non Orthogonal Multiple Access (NOMA), is implemented. 
We propose an analytical model of the wireless cell covered by the BS. 
 Based on this model, we establish a closed form expression of the minimum transmit power of the base station, needed to achieve a given SINR (signal to interference plus noise ratio), whatever the users locations, on the area covered by the base station. 
The closed form expression of the BS transmit power allows to establish quality of service (QoS) and coverage values, in a simple and quick way.


\end{abstract}


\IEEEpeerreviewmaketitle

\section{Introduction}



%

The future 5G wireless network has to answer different challenges such as reducing delays, increasing capacity and throughput of wireless networks.
Moreover, the possibility to connect billions objects receiving data, related to the Internet of Things (IoT), represents also an important issue of the 5G.

In this framework, the Superposition Coding (SC) mechanism may represent an interesting approach. Indeed, it has been established that wireless systems achievable throughput is limited \cite{Shannon48} \cite{Wynner94}. 
One of the interests of superposition coding, also referred to as Non Orthogonal Multiple Access
(NOMA) \cite{Din14}, 
is that SC can allow an achievable throughput closer to Shannon capacity than by using Time Division Multiple Access (TDMA) \cite{GoKe16} \cite{GoTsi14}. \\

The superposition coding approach may also be an interesting approach in the case of IoT, since it is needed to transmit data, simultaneously, to a great number of communicating objects. Several dedicated access networks technologies have been recently proposed for IoT, as
analyzed in \cite{GoGor} \cite{CeVa16}, and also in uplink scenario for IoT \cite{ChaAndre09} \cite{Dhi14}.

An important problem consists in determining the maximal density of
users a base station (BS) can serve, considering transmit, coverage and quality of service constraints,


The paper \cite{Kim2015} shows that the capacity region, using superposition
coding, is achieved in many cases, corresponding to different channel parameter values, by considering a
continuous-time Poisson broadcast channel.

The case of relay-assisted downlink cellular systems is analyzed in \cite{Rini2015}.
This paper shows that, when distributing a message of one user to multiple relays, large gains may be achieved, in terms of rate, provided by superposition
coding and interference decoding.

Analyzing a scenario with two users, one close to a BS and the
other far from it,  the paper \cite{Choi2015} considers  multicast beamforming with superposition coding.
An iterative algorithm is developed, based on an optimal
power allocation, in the aim to find beamforming vectors and powers for both users.
%
%


Assuming gaussian broadcast channels and 
taking into account the receivers power constraints, the paper \cite{Mkim2014} shows
that multi-user transmission schemes, such as superposition
coding are not always optimal in terms of spectral efficiency.

One particular issue considered in \cite{HWang2015} focuses on the
SINR prediction at the scheduler. 
This paper shows that SC provides throughput improvement
for various SINR prediction methods.


The paper \cite{Abbes2016} considers
the application of superposition coding in the aim to increase the system
capacity through multiuser diversity exploitation. 
Authors propose a joint admission control and superposition coding
scheme which allows to provide a good tradeoff between the QoS level
perceived by the user in the system and the utilization of the
scarce radio resources. \\


None of these papers establishes a closed form formula of a base station transmit power in a wireless cell where the superposition coding is implemented. \\

\underline{Our Contribution}: 
We consider a wireless BS and the cell it covers. The superposition coding is implemented. This paper establishes a closed form formula of the BS minimum transmit power needed to reach a SINR target, whatever the location of the user over the area covered by the base station. 
This allows to analyze in a simple and quick way, the quality of service in terms of throughput and spectral efficiency, and the coverage of a BS. \\


%
%

The organization of the paper is as follows. 
In Section \ref{sec:model}, we present the system model. 
In Section \ref{analmodel}, we develop the wireless cell analytical model, which allows to establish the closed form formulas of the transmit power towards a user whatever its location, and the total base station transmit power. 
Section \ref{CoverageQoS} shows that the analysis of coverage and QoS can be done with the closed form formula.
The case of superposition coding without SIC is given is Section \ref{SCsansSIC}.
Section \ref{sec:conclusion} concludes the paper.

\section{System Model} \label{sec:model}

We focus our analysis on the downlink, on the area covered by a base station. The superposition coding is implemented.\\ 
Let remind that superposition coding is also referred to as Non Orthogonal Multiple Access (NOMA) \cite{Din14}, 
Let us consider:

\begin{itemize}

\item ~${N}=\{1,\ldots,N\}$ the set of users connected to the BS, uniformly and regularly distributed over the two-dimensional plane.

\item ~$\gamma^*$ the SINR target to be reached in downlink  by any receiver (user). 


\item ~$r_u$ distance between the transmitting BS and the user $u$.
\item ~${N_u}$ set of users located at a distance from the BS lower or equal to the distance $r_u$.

\item ~${\overline{N}_u}$ set of users located at a distance from the BS higher than $r_u$.


\item ~$P_{u}$ the transmitted power assigned by the base station towards user $u$. 

\item ~$g_{u}$ the propagation gain between the BS and user $u$.

\end{itemize}

Let consider that a target SINR, denoted $\gamma^*$, has to be reached at each user, in the aim to reach a given throughput.
We aim to establish a mathematical expression of the BS transmitting power in the aim to reach this SINR target. 


\subsection{Expression of the SINR with Superposition Coding} \label{SINRSC}

Let consider a BS. N users are located on the area covered by the BS.  
In our analysis, by considering the superposition coding mechanism, we assume that the BS transmits useful power to all the users of the cell it covers, simultaneously, on the same frequency bandwidth. 
We can thus consider the SINR target $\gamma^*(u)$ reached by any user (UE) $u$ connected to the BS, defined by:
\begin{equation} \label{SINRgen}
\gamma^*(u)=\frac{P_{u} g_{u} }{
\sum\limits_{v\in{N},v\neq u}P_{v} g_{u}  + N_{th}},
\end{equation}
where 
(i) $P_{u} g_{u}$ is the useful signal received by $u$

(ii) $\sum\limits_{v\in{N},v\neq u}P_{v} g_{u}$ represents the interference power due to the transmitting powers of the BS towards the other users of the cell, where $P_u$ and  $P_v$ are the transmitting powers towards the mobile $u$ and $v$ and $g_u$ is the pathloss between the BS $b$ and the mobile $u$,  

(iii) and $N_{th}$ is thermal noise power.\\

Considering a pathloss $g_{v}  = K r_v^{-\eta}$, the SINR is expressed as

\begin{equation} \label{SINRpath}
\gamma^*(u)=\frac{P_{u}K r_u^{-\eta} }{
\sum\limits_{v\in{N},v\neq u}P_{v} K r_u^{-\eta}  + N_{th}}.
\end{equation}

\subsection{Successive Interference Cancellation} \label{SupCodSINR}

Since the BS transmits useful powers to all the users at the same time, and on the same frequency, a user $u$ receives a high level of interference due to the powers transmitted towards all others users of the cell. The SIC mechanism allows to mitigate this interference. 
Let rewrite the expression (\ref{SINRpath}) by considering the set of users $N$ as the union of the set $N_u$ of users $v$, such that 
$r_v \leq r_u$ 
and the set $\overline{N}_u$ of users $v$, such that $r_v > r_u$:

\begin{equation} \label{SINRdetail}
\gamma^*(u)=\frac{P_{u}K r_u^{-\eta} }{
\sum\limits_{v\in{N_u},v\neq u}P_{v} K r_u^{-\eta}  + \sum\limits_{v\in\overline{N}_u,v\neq u}P_{v} K r_u^{-\eta}  + N_{th}}.
\end{equation}

For the mobile $u$ to reach a given target SINR $\gamma^*$, it is needed to transmit at a given power $P_u$. 
Due to the pathloss between a transmitter and a receiver, 
mobiles located at distances $r_v > r_u$
(from BS) need a transmitting power 
$P_v > P_u$ in the aim to reach the same SINR $\gamma^*$. 

The successive interference cancellation mechanism allows the user $u$ to decode, using an iterative process, the powers it receives 
dedicated to users located at distances  further than him (from the BS).

Let notice $p^u_u$ the power  
dedicated to user $u$ and received by the mobile $u$ ($p^u_u$= $P_{u} K r_u^{-\eta}$).

Let notice $p^v_u$ the power  
dedicated to user $v$ and received by the mobile $u$ ($p^v_u$= $P_{v} K r_u^{-\eta}$).

Let consider a mobile $u$ needing a useful power $p_u^u$ in the aim to reach the SINR $\gamma^*$. 
Using the SIC mechanism, this mobile $u$ is able to decode any power it receives higher than $p_u^u$. 
And particularly, it is able to decode the power it receives dedicated to the mobile $v$. 
Indeed, since $P_v \geq P_u$, 
the power $p^v_u$ has a higher value than $p^u_u$. 
Therefore, $p^v_u$ can be decoded by the mobile $u$. And thus, it does no more represent an interference for $u$.
This is available for any user $v$ for which  $r_v \geq r_u$.
As a consequence, considering all the users of the cell, in expression (\ref{SINRdetail}), it can be expressed $\sum\limits_{v\in\overline{N}_u,v\neq u}P_{v} K r_u^{-\eta} = 0$, since this term 
does no more represent an interference.
We thus can write the SINR (\ref{SINRdetail}) as:

\begin{equation} \label{SINRSIC2}
\gamma^*(u)=\frac{P_{u} r_u^{-\eta} }{
\sum\limits_{v\in{N_u},v\neq u}P_{v} r_u^{-\eta} + N_{th}/K}.
\end{equation}
which can be written:
\begin{equation} \label{SINRSIC}
\gamma^*(u)=\frac{P_{u} r_u^{-\eta} }{
\sum\limits_{v, r_v \leq r_u}P_{v} r_u^{-\eta} - P_{u} r_u^{-\eta}   + N_{th}/K}.
\end{equation}

From this expression, the total power transmitted to users located at a distance lower or equal to $r_u$ can be expressed as: 

\begin{equation} \label{SINRp}
\sum\limits_{v, r_v \leq r_u}P_{v}= \frac{\gamma^*(u)+ 1}{\gamma^*(u)} P_u - \frac{N_{th}}{K}  r_u^{\eta}.
\end{equation}

\section{Wireless Cell Analytical Model} \label{analmodel}
The analytical model proposed in this paper consists in considering a continuum of users, instead of a discrete set of users, over the cell area, i.e. the area covered by the BS. 
This continuum of users is characterized by a density of users $\rho(r)$.
Let notice that considering a density of users matches well with IoT approaches. \\

\subsection{Base Station Transmit Power} \label{BSPower}
Using this model, it becomes possible to express (\ref{SINRp}) as follows. 
Assuming the cell covered by the BS as a disk of radius $R_c$, we consider a user located at a distance $r$ from its serving BS (dropping the index $u$).
The discrete sum on users $v$ of (\ref{SINRp}) can be expressed as an integral over the cell. 

Considering the SINR target $\gamma^*(r)$, we thus can write:
\begin{equation} \label{SINRp1}
\int_0^{2\pi}\int_{R_0}^r \rho(s) P(s)s ds d\theta  =  \frac{\gamma^*(r)+ 1}{\gamma^*(r)} P(r) - \frac{N_{th}}{K}  r^{\eta} .
\end{equation}
where $P(r)$ is the power transmitted towards the mobile located at distance $r$, $s$ is the variable of integration, $\rho(s)$ is the density of users depending on s 
and $sdsd\theta$ is the element of integration, and $R_0$ the minimum distance between a user and the base station.

A unique SINR target $\gamma^*$ is considered  for all the mobiles of the cell. This corresponds to a service which demands a given throughput for all mobiles. Let assume that the users are uniformly distributed over the area of the cell, $\rho(s)= \rho$ does not depend on $s$. 
Therefore, denoting  $\zeta = \frac{\gamma^*+ 1}{\gamma^*} $, expression (\ref{SINRp1}) can be written as:
\begin{equation} \label{SINRp2}
2\pi \rho \int_{R_0}^r P(s)s ds =  \zeta P(r) - \frac{N_{th}}{K}  r^{\eta}.
\end{equation}

\noindent From expression (\ref{SINRp2}), two important expressions can be derived: 

i) the transmit power towards a user located at distance $r$, 

ii) the total BS transmit power. 

\subsection{Theorem 1: Transmit Power Towards A User} \label{theorem1}
Considering a continuum of users characterized by a density $\rho$, and assuming a unique SINR target $\gamma^*$ over the cell, the transmit power towards a user, denoted $P(r)$, can be expressed as:
\begin{equation} \label{SINRpr2b}
P(r) = b \exp(\frac{ar^2}{2})\times  \int_{R_0}^r s^{\eta-1}\exp(-\frac{as^2}{2})ds ,
\end{equation}
where 
$a= 2 \pi \huge\frac{\rho}{\zeta}$ 
and $b = \frac{N_{th} \eta}{K \zeta}$. 

\noindent The proof is detailed in Appendix 1.

\subsection{Theorem 2: Total BS Transmit Power} \label{theorem2}
Considering a continuum of users characterized by a density $\rho$, and assuming a unique SINR target $\gamma^*$ over the cell, the total power transmitted by the BS, simultaneously  towards all the users of the cell, denoted  $P_{b,sic}$, can be expressed as:
\begin{eqnarray} \label{SINRptot1bis}
P_{b,sic} &=& \frac{N_{th}}{K} R_c^{\eta} \times  \nonumber \\
& &  \left(\eta \exp(\frac{a R_c^2}{2})\times  \int_{R_0/R_c}^{1} t^{\eta-1}\exp(-\frac{a R_c^2}{2}t^2)dt - 1 \right) \nonumber \\
\end{eqnarray}

\noindent The proof is detailed in Appendix 2. \\

Considering $R_0 << R_c$ and thus $\frac{R_0}{R_c} \approx 0 $, and denoting $\beta = \frac{\pi \rho{R_c}^2}{\zeta}$, we can express (\ref{SINRptot1bis}) as:  
\begin{eqnarray} \label{SINRptot21ap2}
P_{b,sic} &=&   \frac{N_{th}}{K} R_c^{\eta} \times  2\beta \exp(\beta) \left( \Gamma(\frac{\eta+2 }{2}, \beta) - \Gamma(\frac{\eta+2}{2})\right) \nonumber \\
\end{eqnarray}
where 
$\Gamma(x) = \int_{0}^{\infty} t^{x-1}\exp(-t) dt$ \\  
\noindent and $\Gamma(u,x) = \int_{x}^{\infty} t^{u-1}\exp(-t) dt$  is the incomplete $\Gamma$ function.

\subsection{Interest of BS transmit power closed form formulas} \label{SINRsansSC}

\begin{figure}[htbp]
\centering
\includegraphics[scale=0.5]{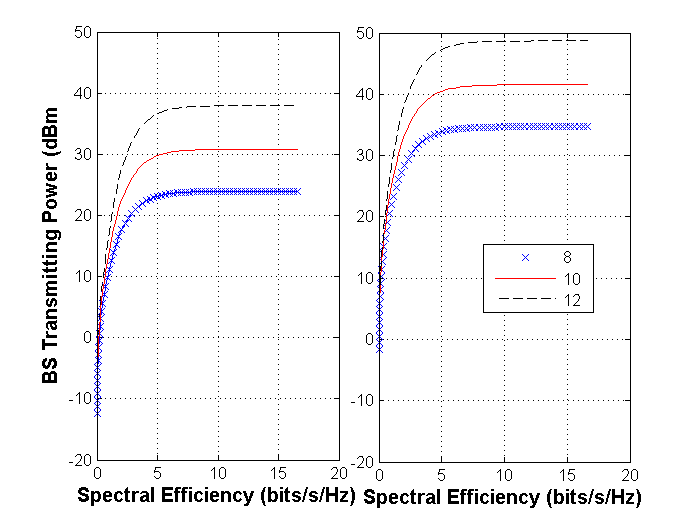}
\caption{\footnotesize Base station Transmitting Power vs Spectral Efficiency for a density of 8, 10 and 12 users per cell area, and for a radius of the zone covered of 50m (left) and 100m (right).}
\label{SICMaPScBaseRc}
\end{figure}

The total BS transmit power closed form formula highlights the impact of the size of the cell $R_c$, the propagation (pathloss factor) $\eta$, the antenna characteristics and the frequency through the parameter $K$, and the thermal noise $N_{th}$. 
It also depends on the density of users and the SINR target value. 
Moreover the analysis of transmit power may be done in a simple and quick way, without simulation, by applying the formula (e.g. fig. \ref{SICMaPScBaseRc}).

Thanks to the SIC mechanism, an important part of interference disappears. Expression (\ref{SINRptot2}) shows that the density of users and the SINR target are, in theory, not limited.
As a consequence, the SINR target and density of users can reach high level values. 
Moreover for a given density of users, the power needed to reach a given value of SINR target reaches a high limit asymptotically, even if the SINR target increases.
This means that for an SINR target higher than a threshold value, an increase of SINR target does not need a high increase of power. 
In the case analyzed in fig. \ref{SICMaPScBaseRc}, this threshold value corresponds to a spectral efficiency of 5 bits/s/Hz.
Let however notice that a high value of the BS transmit power is needed to offer a high SINR target to a high density of users.
Fig. \ref{SICMaPScBaseRc} shows that a transmit power of 41 dBm is needed to reach a spectral efficiency of 15 bits/s/HZ, with a density of 10 users/cell. 
	
	%

\section{Application: Coverage and quality of service} \label{CoverageQoS}
One of the main issue of wireless networks consists in determining the coverage of a BS, and the quality of service it can provide to users.

In the case of superposition coding with SIC the expression of the transmitting power (\ref{SINRptot2}) is particularly interesting since it allows to calculate the area covered by a BS,  in an obvious way. Indeed, expression (\ref{SINRptot2}) highlights the impact of the size of the zone covered, on the transmitting power. The parameter $R_c$ shows that the minimum power needed for a BS to cover a disk of radius R$_c$ is given by expression (\ref{SINRptot2}). In other words the coverage can be directly calculated by using (\ref{SINRptot2}), for a given transmitting power.

  
Moreover, expression (\ref{SINRptot2}) also highlights the quality of service, in terms of throughput or spectral efficiency. Indeed, the SINR target $\gamma^*$ allows calculating the spectral efficiency (\ref{SE})  and the throughput (\ref{Throughput}) that can be reached by users. 
In other words the QoS can directly be calculated, for a given transmitting power, by using expression (\ref{SINRptot2}).

 Fig \ref{SICMaPScBaseRc} (left side) shows that a transmitting power of 24 dBm is needed to cover an area of radius R$_c$=50m in the aim to reach a spectral efficiency of 5 bits/s/Hz, for a density of 8 users/cell area (K = 2.66. $10^{-4}$ and $\eta$ = 3.57  \cite{3GPP2012}).
When the density of users increases, the transmitting power increases, too. 
When the radius area $R_c$ increases, the needed transmitting power increases, too (Fig \ref{SICMaPScBaseRc}, right side).
This figure  also shows that over a given spectral efficiency (about 5 bits/s/Hz), the transmitting power increases very slowly towards an asymptotic value, given by (\ref{SINRptot2}) when $\gamma^*$ $\rightarrow \infty$ (i.e. when $\zeta$ $\rightarrow$ 1), and expressed  as:

\begin{eqnarray} \label{SINRptot3}
P_{b,sic} &=& \frac{N_{th}}{K} R_c^{\eta} \left( \eta \exp(\beta_2) \int_{0}^{1} t^{\eta-1}\exp(-\beta_2 t^2)dt - 1 \right) \nonumber \\
\end{eqnarray}
where $\beta_2 = \pi \rho{R_c}^2$

Let remind that, for a given SINR $\gamma$, the spectral efficiency SE is given by 
\begin{eqnarray} \label{SE}
 SE = \log_2(1+ \gamma)
\end{eqnarray}
And the throughput D is given by 

\begin{eqnarray} \label{Throughput}
 D = W \log_2(1+ \gamma)
\end{eqnarray}
where W is the frequency bandwidth.

\section{Superposition Coding without SIC} \label{SCsansSIC}

It is interesting to express the BS transmitting power, when the SIC mechanism is not implemented.
Since in this case all users of the cell contribute to interference except $u$, the equation (\ref{SINRp}) becomes 
\begin{equation} \label{SINRpsans}
\sum\limits_{v, r_v \leq R_c}P_{v}  = \frac{\gamma^*(u)+ 1}{\gamma^*(u)} P_u - \frac{N_{th}}{K}  r_u^{\eta}.
\end{equation}
Considering a uniform density of users $\rho$ independent of r, and a unique SINR target $\gamma^*$ for all users, we can write:
\begin{equation} \label{SINRptot1}
\int_0^{2\pi}\int_{R_0}^{R_c} \rho P(s)s ds d\theta  = \zeta P(r) - \frac{N_{th}}{K}  r^{\eta} 
\end{equation}
Let notice that, in this case, the integral does not depend on r. Since the total transmitting power without SIC, denoted $ P_{BS}$ can be expressed as  $ P_{BS} = \int_0^{2\pi}\int_{R_0}^{R_c} \rho P(s)s ds d\theta $, (\ref{SINRptot1}) can be expressed as:

\begin{equation} \label{SINRptotsans2}
\zeta P(r) = P_{BS} + \frac{N_{th}}{K}  r^{\eta}
\end{equation}
so we have, by integration over the cell:
\begin{eqnarray} \label{SINRptotsans3}
\zeta \int_0^{2\pi}\int_{R_0}^{R_c} \rho P(s)s ds d\theta &=& \int_0^{2\pi}\int_{R_0}^{R_c} \rho P_{BS} s ds d\theta \nonumber \\ 
&+&   \int_0^{2\pi}\int_{R_0}^{R_c} \rho  \frac{N_{th}}{K}  s^{\eta} s ds d\theta \nonumber \\
\end{eqnarray}

Therefore, the total BS transmitting power $P_{BS}$  without SIC, can be expressed as: 

\begin{eqnarray} \label{SINRptot2sans5}
P_{BS} &=& \frac{2\pi \rho \frac{N_{th}}{K} {R_c}^{\eta+2}(1- (R_0/{R_c})^{\eta+2})/(\eta+2)}{\zeta - \pi \rho {R_c}^2 (1-({R_0}/{R_c})^{2})}  \nonumber \\
\end{eqnarray}
Considering $R_0 << R_c$ and thus $\frac{R_0}{R_c} \approx 0$ : 
\begin{eqnarray} \label{SINRptot2sans6}
P_{BS} &=& \frac{2\pi \rho \frac{N_{th}}{K} {R_c}^{\eta+2}/(\eta+2)}{\zeta - \pi \rho{R_c}^2}  \nonumber \\
\end{eqnarray}
and

\begin{eqnarray} \label{SINRptot2sans7}
P_{BS} &=& \frac{2 \frac{N_{th}}{K} {R_c}^{\eta}/(\eta+2)}{  \frac{\zeta}{\pi \rho{R_c}^2}   - 1 } 
\end{eqnarray}

%
%
%

\section{Conclusion} \label{sec:conclusion} 
We established closed form expressions of the transmit power needed to achieve a given SINR, whatever users locations in a wireless cell, by considering the Superposition Coding and SIC mechanism. 
We showed that the analysis of the quality of service and the coverage of a base station can be done in a simple and quick way, by using the expression of the total BS transmit power.
In this paper, we aimed to highlight the specific impact of the superposition coding mechanism. Therefore, the analysis was focused on a unique wireless cell.
In a future work, the analysis will be extended to the case of a network, constituted by a great number of base stations, and will take into account the interference they induce.

\section*{Appendix 1: Transmit Power Towards a User} \label{SINRRRH}
The key method to establish an expression of the transmit power consists in derivating expression (\ref{SINRp2}) with respect to r. It thus can be written:

\begin{equation} \label{SINRpr1}
2\pi \rho P(r)r =  \zeta P'(r) - \frac{N_{th}}{K} \eta r^{\eta-1}.
\end{equation}
where $P'$ means the derivative of $P$ according to $r$.
This is a differential equation with non-constant coefficients. The solution allows to express 
the transmitting power towards a user located at a distance $r$ from its serving BS as: 
\begin{equation} \label{SINRpr2b}
P(r) = b \exp(\frac{ar^2}{2})\times  \int_{R_0}^r s^{\eta-1}\exp(-\frac{as^2}{2})ds ,
\end{equation}
where 
$a= 2 \pi \huge\frac{\rho}{\zeta}$ 
and $b = \frac{N_{th} \eta}{K \zeta}$. 

\section*{Appendix 2: Total BS Transmit Power} \label{SINRRRH}

The total transmitting power of the BS, $P_{b,sic} $, can be expressed as: 
\begin{equation} \label{SINRptot0}
P_{b,sic} = \int_0^{2\pi}\int_{R_0}^{R_c} \rho(s) P(s)s ds d\theta = 2 \pi \rho \int_{R_0}^{R_c} P(s)s ds d\theta 
\end{equation}

From (\ref{SINRp2}) and (\ref{SINRpr2b}), by integrating until $r = R_c$, it can be expressed as:

\begin{equation} \label{SINRptot}
P_{b,sic} = \int_0^{2\pi}\int_{R_0}^{R_c} \rho(s) P(s)s ds d\theta 
= \zeta P(R_c) - \frac{N_{th}}{K} R_c^{\eta}.
\end{equation}
and thus:
\begin{eqnarray} \label{SINRptot1}
P_{b,sic} &=& \int_0^{2\pi}\int_{R_0}^{R_c} \rho(s) P(s)s ds d\theta \nonumber \\
&=& \zeta P(R_c) - \frac{N_{th}}{K} R_c^{\eta} \nonumber \\
&=& \zeta b \exp(\frac{a R_c^2}{2})\times  \int_{R_0}^{R_c} s^{\eta-1}\exp(-\frac{as^2}{2})ds - \frac{N_{th}}{K} R_c^{\eta} \nonumber \\
&=& \frac{N_{th} \eta}{K} \exp(\frac{a R_c^2}{2})\times  \int_{R_0}^{R_c} s^{\eta-1}\exp(-\frac{as^2}{2})ds - \frac{N_{th}}{K} R_c^{\eta} \nonumber \\
&=& \frac{N_{th}}{K} R_c^{\eta} \times  \nonumber \\
& &  \left( \eta \exp(\frac{a R_c^2}{2})\times  \int_{R_0/R_c}^{1} t^{\eta-1}\exp(-\frac{a R_c^2}{2}t^2)dt - 1 \right) \nonumber \\
\end{eqnarray}

Considering $R_0 << R_c$ and thus $\frac{R_0}{R_c} \approx 0$ : 

\begin{eqnarray} \label{SINRptot2}
P_{b,sic} &=& \frac{N_{th}}{K} R_c^{\eta} \times \nonumber \\
& & \left( \eta \exp(\frac{\pi \rho{R_c}^2}{\zeta}) \int_{0}^{1} t^{\eta-1}\exp(-\frac{\pi \rho{R_c}^2}{\zeta}t^2)dt - 1 \right)  \nonumber \\
\end{eqnarray}

Using $\Gamma$ function, we notice that expression can also be written as:  
\begin{eqnarray} \label{SINRptot21ap2}
P_{b,sic} &=&   \frac{N_{th}}{K} R_c^{\eta} \times  2\beta \exp(\beta) \left( \Gamma(\frac{\eta+2 }{2}, \beta) - \Gamma(\frac{\eta+2}{2})\right) \nonumber \\
\end{eqnarray}
where $\Gamma(x) = \int_{0}^{\infty} t^{x-1}\exp(-t) dt$   

\noindent and $\Gamma(a,x) = \int_{x}^{\infty} t^{a-1}\exp(-t) dt$  is the incomplete $\Gamma$ function.

\end{document}